# Real Time and Energy Efficient Transport Protocol for Wireless Sensor Networks

**S.Ganesh**
Research Scholar, Sathyabama University, Chennai-600119
Email: ganesh_smit@yahoo.com
**Dr.R.Amutha**
Department of Electronics and Communication Engineering,
Sri Venkateswara College of Engineering Sriperumbudhar-602105
Email: ramutha@svce.ac.in

---------------------------------------------------ABSTRACT-------------------------------------------------
Reliable transport protocols such as TCP are tuned to perform well in traditional networks where packet losses occur mostly because of congestion. Many applications of wireless sensor networks are useful only when connected to an external network. Previous research on transport layer protocols for sensor networks has focused on designing protocols specifically targeted for sensor networks. The deployment of TCP/IP in sensor networks would, however, enable direct connection between the sensor network and external TCP/IP networks. In this paper we focus on the performance of TCP in the context of wireless sensor networks. TCP is known to exhibit poor performance in wireless environments, both in terms of throughput and energy efficiency. To overcome these problems we introduce a mechanism called TCP Segment Caching .We show by simulation that TCP Segment Caching significantly improves TCP Performance so that TCP can be useful even in wireless sensor

## 1. INTRODUCTION

Wireless sensor networks are composed of a large number of radio-equipped sensor devices that autonomously form networks through which sensor data is transported. The devices are typically severely resource-constrained in terms of energy, processing power, memory, and communication bandwidth. Many applications of wireless sensor networks require an external connection to monitoring and controlling entities that consume sensor data and interact with the sensor devices. Running TCP/IP in the sensor network makes it possible to connect the sensor network directly to IP-based network infrastructures without proxies or middle-boxes. Since each sensor device is able to communicate using TCP/IP, [1] it is possible to route data to and from the sensor network using standard IP-based technologies such as General Packet Radio Service (GPRS).

Data transport in IP-based sensor networks is performed using the two main transport protocols in the TCP/IP stack: The best-effort UDP and the reliable byte-stream protocol TCP. UDP is used for sensor data and other information that do not use unicast reliable byte-stream transmission. TCP should be used for administrative tasks that require reliability and compatibility with existing application protocols.

## 2. DRAWBACKS OF TCP

TCP is a connection-oriented protocol. Before data transmission, there is a three-way handshake interactive process. The three-way handshake process will be a big overhead for the small volume data. Also since wireless link is error-prone under WSNs, the time to setup TCP connection might be much longer than that under Internet. Then the data will be probably outdated after TCP connection has been established.

In TCP, it is assumed that all segment losses are resulted from congestion and will trigger window-based flow control and congestion control. This style will incur that TCP will unwisely reduce transmission rate under WSNs when there is no congestion, but packet losses from bit-error. The behavior will lead to low throughput especially under multiple wireless hops.

TCP uses end-to-end approach to control congestion. This approach generally has longer response time when congestion occurs, and in-turn will result in lots of segment dropping. The segment dropping means useless energy consumption and not energy-efficient.

TCP uses end-to-end ACK and retransmission to guarantee reliability. This approach will cause much lower throughput and longer transmission time if Round-Trip Time is larger as that in large-scale WSNs, since the sender will stop to wait for the ACK after each data transmission.

Under WSNs, sensor nodes may have different hops and different RTT from sink. TCP in such environment may cause unfairness. The sensor nodes near to sink may get more opportunities to transmit data and may deplete their energy first, and the whole wireless sensor network will be disjointed with a high probability

**3. IP-BASED SENSOR NETWORKS**

Besides poor [2] TCP performance both in terms of throughput and energy-efficiency, there are other problems with TCP/IP that must be solved before IP-based sensor networks can become ubiquitous. In this section we identify these problems and sketch solutions.

**Address centric addressing:** The IP addresses in traditional IP networks are assigned to each network interface based on the network topology. Each network interface is assigned a unique IP address using either manual configuration or semi-automated mechanisms such as DHCP. Such address assignment mechanisms are not suited for large scale sensor networks. Instead, IP-based sensor networks may perform spatial IP address assignment that uses the spatial location of the sensor nodes to construct semi-unique IP addresses.

**Address centric routing:** In traditional IP networks, each packet is transparently routed through the network. The routing path is based on the IP addresses and the topology of the network. For wireless sensor networks, data centric routing mechanisms are often preferable [3]. To implement data centric routing in IP-based sensor networks, we use application overlay networks.

**Header overhead:** Compared to specialized sensor networking protocols, the protocols in the TCP/IP suite have a very large header overhead. The shared context nature of sensor networks enables efficient header compression to reduce TCP/IP header overhead.

**4. TCP SEGMENT CACHING**

The key idea of TCP Segment Caching is to avoid energy-costly end-to-end retransmissions by[4]caching TCP segments inside the network and retransmitting segments locally, i.e. from the intermediate sensor nodes' caches, when packet loss occurs. Ideally, each node would cache all segments and perform the retransmission exactly from the last node that has transmitted a segment before it has been lost. However, due to the constrained resources of the sensor nodes, we assume that each node can only cache one segment. Nodes take extra care to cache segments presumably not received by the next node. The Segment caching is only implemented in the intermediate sensor nodes and does not require any changes on the TCP endpoints. A sensor node acting as the receiver may make use of the following standard TCP features: The receiver announces a small maximum segment size in order to avoid large TCP segments exceeding the capacity of the sensor nodes. Further, the receiver announces a small window size constraining the number of segments in flight.

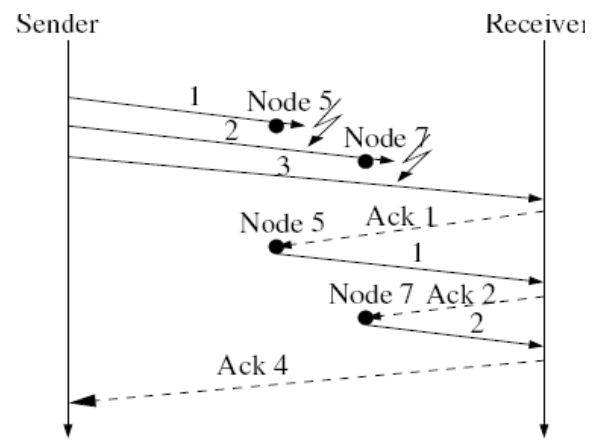

Figure1. TCP Segment Caching

Figure 1 shows a simplified example of DTC. To keep the example simple, we assume that nodes are able to detect when a TCP segment they have transmitted is lost. In this example, a TCP sender transmits three TCP segments. Segment 1 is cached by node 5 right before it is dropped in the network, and segment 2 is cached by node 7 before being dropped. When receiving segment 3, the TCP receiver sends an acknowledgment (ACK 1).

We assume here that node 7 must not retransmit segment 2 when it receives ACK 1 since this acknowledgment comes too early. When receiving ACK 1, node 5, which has a cached copy of segment 1, performs a local retransmission. Node 5 also refrains from forwarding the acknowledgment towards the TCP sender, so that the acknowledgment segment does not have to travel all the way through the network. When receiving the retransmitted segment 1, the TCP receiver acknowledges this segment by transmitting ACK 2. On reception of ACK 2, Node 7 performs a local retransmission of segment 2, which was previously cached. This way, the TCP receiver obtains the two dropped segments by local retransmissions from sensor nodes in the network, without requiring retransmissions from the TCP sender. When the acknowledgment ACK 4 is forwarded towards the TCP sender, sensor nodes on the way can clear their caches and are thus ready to cache new TCP segments.

### 4.1 Packet Loss Detection

To avoid end-to-end retransmissions, [5] [6] TCP segment caching needs to respond faster to packet loss than regular TCP. Segment caching relies mainly on timeouts to detect packet loss. Every node participating in Segment caching maintains a soft TCP state for connections that pass through the node. We assume symmetric and relatively stable routes, and therefore the nodes can estimate the delays between the node and the connection end-points. Each node measures the round-trip time (rtt) to the receiver and adapts a retransmission timeout to 1.5 * rtt. This ensures that the retransmission timeout values are smaller for nodes close to the destination and higher for nodes close to the source.

Since the rtt values experienced by the nodes are lower than those estimated by the TCP end-points, the intermediate nodes are able to perform retransmissions earlier than the TCP end-points. Segment caching nodes set a timer for a local retransmission when they lock a segment in the cache. Simulations have shown that the other standard TCP mechanism to detect packet loss, duplicate acknowledgements, cannot contribute significantly to the performance of DTC.

### 4.2 Selective Acknowledgements

The Segment caching mechanism uses the TCP (Selective Acknowledgement) SACK [7] option to both detect packet loss and as a signaling mechanism between sensor nodes. TCP segment caching uses the latter to inform other nodes about the segments locked in the cache. On reception of a TCP ACK with an acknowledgement number smaller than the sequence number of its cached segment a node performs the following actions:

• If a node's cached segment's sequence number (cached) is not in the SACK block, the node retransmits the cached segment. Before transmitting the TCP ACK towards the sender, the node adds cached to the SACK block. Moreover, if cached fills all gaps, i.e. with cached all segment numbers up to the highest in the SACK block are acknowledged, the node can drop the acknowledgement. Note that the node should not generate a new ACK acknowledging all the segments in the SACK blocks since the receiver is allowed to discard a previously selective acknowledged segment.
• The node can clear its cache if the cached segment's sequence number is in the SACK block since this means that either the receiver has received the corresponding segment or that the segment is cached and locked by a node closer to the receiver.
• While TCP Segment caching caches TCP data segments, it does not cache and retransmit TCP ACKs. The segment caching uses a simple local regeneration of TCP acknowledgements. When an intermediate node sees a TCP data segment, for which it has already received and forwarded a TCP ACK, the node assumes that the TCP ACK has been lost. Therefore, it does not forward the data segment but instead locally regenerates a TCP ACK.

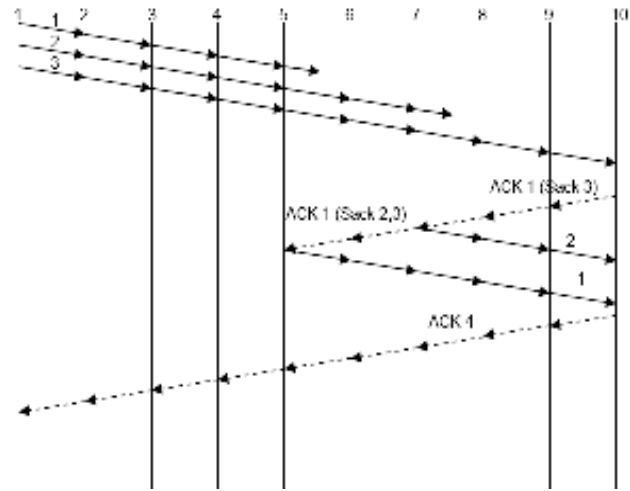

Figure2. Segment caching with SACK

Figure 2 shows an example of TCP segment caching using SACK as a signaling mechanism. In this example, a TCP sender transmits three TCP segments. Segment 1 is cached by node 5 before it is dropped in the network. Since node 5 does not receive a link layer ACK, it locks segment 1 in the cache. Similar, segment 2 is cached and locked by node 7. When receiving segment 3, the receiver sends an acknowledgment ACK 1 with a SACK block for segment 3.

On reception of the ACK segment, node 7 retransmits segment 2, adds a SACK block for segment 2 and forwards the acknowledgment. Eventually node 5 receives that acknowledgment and retransmits segment 1. Since all the gaps are filled now, node 5 drops the acknowledgment. Having received both segment 1 and segment 2, the receiver transmits ACK 4.

A selective acknowledgment option indicates either that the receiver has received an out-of-order segment, or that a sensor node closer to the receiver has locked the segment in its cache. A Sensor node that sees a selective acknowledgment for a segment it has in its cache can therefore clear the cache.

### 5. RESULTS

We have implemented TCP Segment caching and performed evaluations in the OMNet++ discrete event simulator [8]. We have performed simulations with unidirectional TCP data transfers, with and without the DTC mechanism enabled. The data transfers consisted of 500 TCP segments. We use a chain topology as shown in Figure 3.

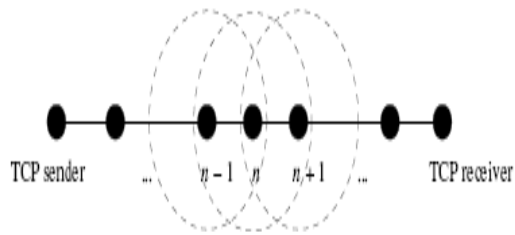

Figure 3. Simulation Topology

Where node n is in transmission range of node n – 1 and n + 1, but node n -1 is not in range of node n + 1. We performed the simulations until the sender receives the acknowledgment for the 500th segment. For our simulations, we have implemented a link layer with explicit positive link layer acknowledgments. Since TCP data segments are larger than acknowledgments, we set the packet loss probability for data segments to twice the loss probability for TCP acknowledgments and to four times the loss probability of link level acknowledgments.

We use a uniformly distributed packet loss model. Our simulations consist of 30 runs, and the reported results are the average of the 30 runs. The results indicate that DTC brings vast improvements: For path lengths between 6 and 11 hops and per-hop packet loss rates between 5% and 15%1, the number of end-to-end retransmissions performed by the sender decreases by a factor of ten. The amount of end-to-end retransmissions decrease even more for higher packet loss rates and longer paths.

5.1 Load Reduction near the Sender

In sensor networks, sensor data flows from nodes that collect sensor data to sinks, whereas control or management data flows from sinks to sensor nodes Therefore, nodes close to the sink usually are the first to run out of energy because sensor data has to be routed through them. Thus, a transport protocol should shift the burden [9] from these nodes to nodes inside the network. Performing local retransmissions instead of end-to-end retransmissions could obviously assist in that task. In our simulations we have counted the number of transmissions of data segments each node has to perform.

Figure 4 shows the results for 11 hops and a packet loss rate of 10% for data packets. In the figure, the numbers on the y-axis denote the average number of transmissions per run and node. In the figure, node 0 is the node closest to the TCP sender (sensor data sink), node 9 the node closest to the TCP receiver. The figure shows that without Segment caching, nodes close to the sink have to transmit much more segments than nodes further away from the sink. With segment caching, load is reduced at nodes close to the sink and evenly distributed among the nodes on the path.
In fact, using segment caching, the vulnerable nodes close to the sink perform slightly less transmissions than nodes close to the receiver.

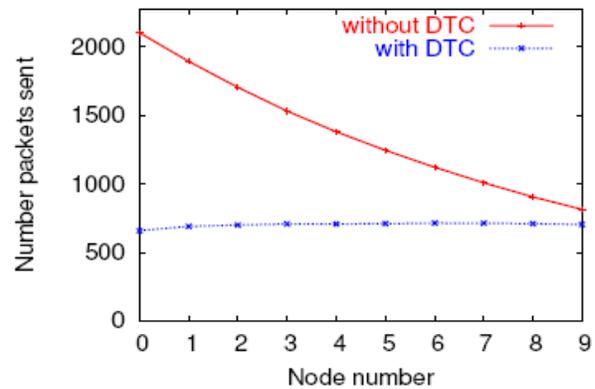

Figure 4. Load Reduction near the sender

5.2 Throughput

In wireless sensor networks with low communication bandwidth, resource-constrained nodes, and high packet loss rates, TCP throughput cannot be expected to be high. We expect that TCP segment caching, by performing local retransmissions, can increase the TCP throughput

6. CONCLUSION

In this paper, we have presented TCP Segment Caching. It enhances TCP performance in sensor networks both in terms of energy efficiency and throughput. The TCP Segment caching achieves this by caching TCP segments inside the sensor network and retransmitting lost segments locally. Furthermore, the segment caching shifts the burden of the load from vulnerable nodes close to the base station into the sensor network. There are more ideas and trade-offs to be explored.

The future work may explore the following Possibilities:

An Adaptive congestion control that integrates end-to-end and hop-by-hop may be more helpful for wireless sensor networks with diverse applications on it, and useful for energy-conservation and simplification of sensor operation

An Adaptive recovery mechanism is required to support packet-level and application-level reliability, and to be helpful for energy-conservation.

Existing transport control protocols have hardly implemented any cross-layer optimization. However lower-layers such as network layer [10] and MAC layer can provide useful information up to transport layer. A new effective and cross-layer optimized transport control protocol can be available through such cross-layer optimization.

For some application, WSNs only needs to correctly receive packets from a certain area and not from every sensor nodes in this area, or some ratio of successful transmission from a sensor node will be sufficient. These new reliability can be utilized to design more efficient transport control protocols

## Authors Biography

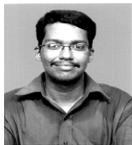

**S.Ganesh** received the B.E., degree in Electronics and communications Engineering from Bharadidasan University Trichy, Tamilnadu ,India and M.E degree in Applied Electronics from Anna University, Chennai, Tamilnadu, India, in 1998, and 2008 respectively. He is working towards his Ph.D. in the area of **Efficient Transport Protocol for Wireless Sensor Networks** as a part-time candidate in ,**Sathyabama University,Chennai**, Tamilnadu, India. He is working as a Senior Lecturer in the department of Electronics and Communication Engineering at **Panimalar Institute of Technology, Chennai**. He has presented papers at IEEE International Conferences in Kerala, Allahabad and in China. He has completed Eleven years in the field of Teaching.

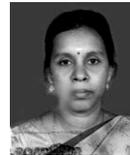

**Dr.R.Amutha** received the B.E., degree in Electronics and communications engineering from Madurai Kamaraj University, Tamilnadu, India and M.E degree in Applied Electronics from PSG College of Technology, Coimbatore, Bharathiar University Tamilnadu, India, in 1987, and 1991 respectively, and the Ph.D. degree in **Communication Networks** from Anna University, Chennai, India in 2006. She is presently working as a professor in Electronics and Communication Engineering, **Sri Venkateswara College of Engineering, Sriperumbudur**, Tamilnadu, India .She served as a Lecturer from June 1988 to May 1993 in the Dept. of ECE, PSG College of Technology, Coimbatore, India. She has presented papers at IEEE International conferences several times.  She has Published papers in International Journals like "International Journal on Information Sciences" and "International Journal on Electronics and Telecommunication Research". Presently she is coordinating the Central Government sponsored research project titled "Artificial Intelligence based tsunami". She has completed Twenty One years in the field of Teaching.